\documentclass{article}
\usepackage{waspaa19,amsmath,graphicx,url,times}
\usepackage{color}

\usepackage{amsmath}
\usepackage{multirow, boldline}
\usepackage[most]{tcolorbox}

\title{Dilated FCN: Listening Longer to Hear Better}

\name{Shuyu Gong$^{1}$, Zhewei Wang$^{1}$, Tao Sun$^{1}$, Yuanhang
  Zhang$^{1}$, Charles D. Smith$^{2}$, Li Xu$^{3}$, Jundong Liu$^{1}$}
\address{$^1$ School of EECS, Ohio University, Athens OH 45701 \\
  $^2$ Department of Neurology, University of Kentucky, Lexington KY 40506 \\
  $^3$ Department of Communication Disorders, Ohio University, Athens
  OH 45701 }

\begin{document}

\ninept
\maketitle


\begin{abstract}

  Deep neural network solutions have emerged as a new and powerful
  paradigm for speech enhancement (SE). The capabilities to capture
  long context and extract multi-scale patterns are crucial to design
  effective SE networks.  Such capabilities, however, are often in
  conflict with the goal of maintaining compact networks
  to ensure good system generalization.

  In this paper, we explore dilation operations and apply them to
  fully convolutional networks (FCNs) to address this issue. Dilations
  equip the networks with greatly expanded receptive fields, without
  increasing the number of parameters. Different strategies to fuse
  multi-scale dilations, as well as to install the dilation modules
  are explored in this work. Using Noisy VCTK and AzBio sentences
  datasets, we demonstrate that the proposed dilation models
  significantly improve over the baseline FCN and outperform the
  state-of-the-art SE solutions.
  
\end{abstract}


\section{Introduction}

Enhancement of audio signals in noisy environments plays an important
role in many speech-related applications, such as 
speech recognition, hearing aids, and cochlear implants. Traditional
speech enhancement (SE) techniques commonly operate on the spectral
domain and rely on certain high-level features to identify target
audio patterns for noise reduction.  Spectral subtraction,
spectral amplitude fitting, 
Wiener filtering, 
and non-negative matrix factorization 
are among the operations that have been extensively studied. 

In recent years, deep neural network-based models (DNNs) have emerged
as a new and more powerful paradigm for many artificial intelligence
(AI) related applications, including speech enhancement. Unlike
traditional machine learning approaches, where certain hand-crafted
features (such as fundamental frequency, formants, MFCC, etc.) need to
be defined and extracted, DNN models carry out feature extraction in
an automatic, data-driven fashion, greatly simplifying the system
design. DNN models also facilitate a common platform for the solutions
across different application areas, including computer vision and
speech signal processing, to be effectively shared.  Up to
date, 
the DNN models that have been explored for speech enhancement include
autoencoder (AE) \cite{Lu_2013_IS}, restricted Boltzmann machine (RBM)
\cite{Wang_2012_IS,Wang_2012_NIPS,Xu_2014_T}, multilayer perceptron
(MLP) \cite{Wang_2014_T}, convolutional neural networks (CNN)
\cite{Hui_2015_C}, recurrent neural networks (RNN) \cite{Tan_2019_T},
generative adversarial network (GAN) \cite{pascual2017segan}, and
fully convolutional networks (FCN) \cite{Tsao_2017_C,Tsao_2018_T},
among others.

Many of the existing SE networks operate on certain time-frequency
(T-F) representation of audio signals, generated through short-time
Fourier transform (STFT) on fixed-length frames. T-F representations
bring great convenience to 
directly target on the frequency components of the audio signals.  The
transformations from the waveform inputs to T-F representations and
back to the waveform outputs, however, impose a structural constraint
to the networks, which complicates the system design and makes it
difficult to predict the network performance.

FCNs on waveform provide a handy and powerful alternative. FCN models
were originally developed as image segmentation solutions
\cite{long2015fully, ronneberger2015u, chen_isbi_2017, wang_mlmi_2018}
and have since been successfully adopted for image modality conversion,
super-resolution and speech signal denoising \cite{Tsao_2017_C,
  Tsao_2018_T}.  The fundamental goal of FCNs is to find mappings,
with certain desired property, between paired signal sources;
therefore they are well-suited to extracting clean waveforms out of
noisy inputs \cite{Tsao_2017_C, Tsao_2018_T}.
The success of
FCNs, 
in great part, is due to their capability of processing input data
from multiple spatial or temporal scales. Further improvements over
the existing FCNs
can be pushed forward by ensuring the models to capture longer
contextual information and/or to enhance multi-scale processing.  The
former (longer context) can be easily achieved through the utilization
of larger filters, while making the network deeper provides a solution
for the latter.  However, a simple combination of these two strategies
will lead to a significantly increased number of parameters, which
would potentially result in limited system generalization and poor
performance in handling diverse noise conditions.

In this paper, we address this challenge by exploring dilated
convolution operation and applying it to FCNs. Dilated convolution was
originally developed for image segmentation, and its effectiveness to
the task has been demonstrated in a number of works
\cite{YuKoltun2016, chen2018deeplab}. We propose to adopt this
operation to improve SE FCNs, by allowing the networks to capture
longer contexts, a.k.a., ``listen longer'', as well as to extract more
diverse, multi-scale audio features. These 
features are then combined through a {\it pyramid pooling} strategy.
We also explore different network locations to embed the new
operations. All the improvements are made without incorporating
additional layers or parameters.
Using Noisy VCTK \cite{valentini2016investigating,valentini2017noisy}
and AzBio sentences \cite{spahr2012development} datasets, we are able
to demonstrate our proposed dilated FCNs outperform the
state-of-the-art solutions.


\section{Method}

Our goal is to develop an 
FCN addition to advance the state-of-the-art of speech enhancement.
The design is based the consideration of allowing the networks to
listen longer 
without an increase of the number of parameters. Our efforts are
focused on the explorations of
1) dilation convolutions to enlarge the receptive fields of neurons;
2) atrous spatial pyramid pooling (ASPP) module to fuse the
multi-scale feature maps;
and 3) 
different network locations in the baseline FCN to embed the ASPP
modules.



\subsection{Baseline FCN model}



Our baseline FCN model is adopted and modified from U-Net
\cite{ronneberger2015u}, which was originally designed for
segmentation of cell images. U-Net has an encoder-decoder
architecture: in the encoding path, input images are processed through
a number of convolution + pooling layers to generate high-level latent
features, which are then progressively upsampled along the decoding
path to reconstruct the target pixel labels.

\begin{figure}[htb]
  \begin{minipage}[b]{1.0\linewidth}
  \centering
  \centerline{\includegraphics[width = 8.5cm]{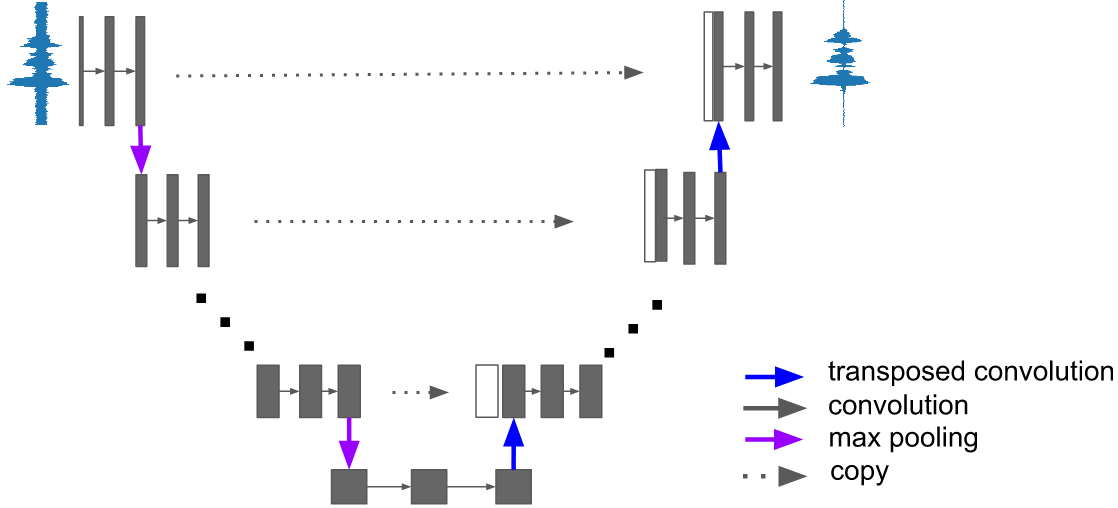}}
\end{minipage}
\caption{Network structure of our Speech-U-Net. Picture is best viewed
  in color. }
\label{fig:res-seg}
\end{figure}

To fit our data and task, we modified the original U-Net as
follows. First, as the inputs and outputs of our model are
one-dimensional waveforms, we replace all the two-dimensional
convolution operations in U-Net with one-dimensional convolutions. We
keep the original U-Net structure of two convolution layers followed
by one pooling/upsampling layer. 
The network is made deeper to enhance its capability to capture the
features in more scales. The encoding path of our modified U-Net now
has 6 convolution-pooling blocks of totally 18 layers.
We also use padding in convolution/deconvolution layers to maintain
the spatial dimension so that the skip connections can directly
concatenate encoding layers with the corresponding decoding layers.
The $L_1$ distance between network predictions (noise-reduced speech)
and the ground-truth (clean speech) is used as the objective
function. We term our baseline model Speech-U-Net, whose structure is
shown in Fig.~\ref{fig:res-seg}.

\subsection{Dilation module}
Noisy speech signals tend to contain components with 
diverse frequency profiles.  
To capture them through convolutions,
filters of varying sizes would be required. Small filters work well in
catching high-frequency noise, but not so effectively for
low-frequency sounds. Large filters perform in an opposite way, keen
to extract low-frequency components but not high-frequency noise.

The receptive field (RF) at each layer decides the length of the audio
signal a neuron can hear. In convolution-based neural networks,
including FCNs, the RFs are increased along layers through both
pooling and
convolutions. Let $R_k$ be the RF of neurons on layer $l_k$, and it
can be computed as:
\begin{equation} \label{eq:cnn}
  R_k=R_{k-1}+((f_k-1)\times\prod_{i=1}^{k-1}s_i)
\end{equation}
where $f_i$ and $s_i$ are the sizes and strides of the filters on
layer $l_i$, respectively. It should be noted that a max pooling of
$n \times 1$, from the RF perspective, has the same effect of
convolutions of stride equal to $n$, and the standard convolutions
(stride equal to 1) increase the RFs in a linear fashion.

The neurons at the final layers of a network may have RFs that have
been enlarged multiple times (e.g., $2^5 = 32$ after 5 max pooling
operations).
However, due to the high sampling rates of audio signals, the feature
maps in speech-enhancement FCNs tend to catch speech segments with
short durations.  In our baseline model, which has 5 pooling layers
and filters of size 30 at each convolution layer, each neuron at the
last layer of the encoding path covers 3686 sample points from the
input, which is a 0.23 second sound clip sampled at the rate of
$16,000$ per second (only $0.077$ second for the rate of $48,000$ per
second), quite insufficient to grasp all types of audio patterns.  To
gain long enough contextual information, one can simply add more
pooling layers to makes the network deeper, but that would inevitably
lead to a more complicated system with an increased number of
parameters, as well as longer training and inference times.

\begin{figure}[htb]
  \begin{minipage}[b]{1.0\linewidth}
  \centering
  \centerline{\includegraphics[width = 7.5cm]{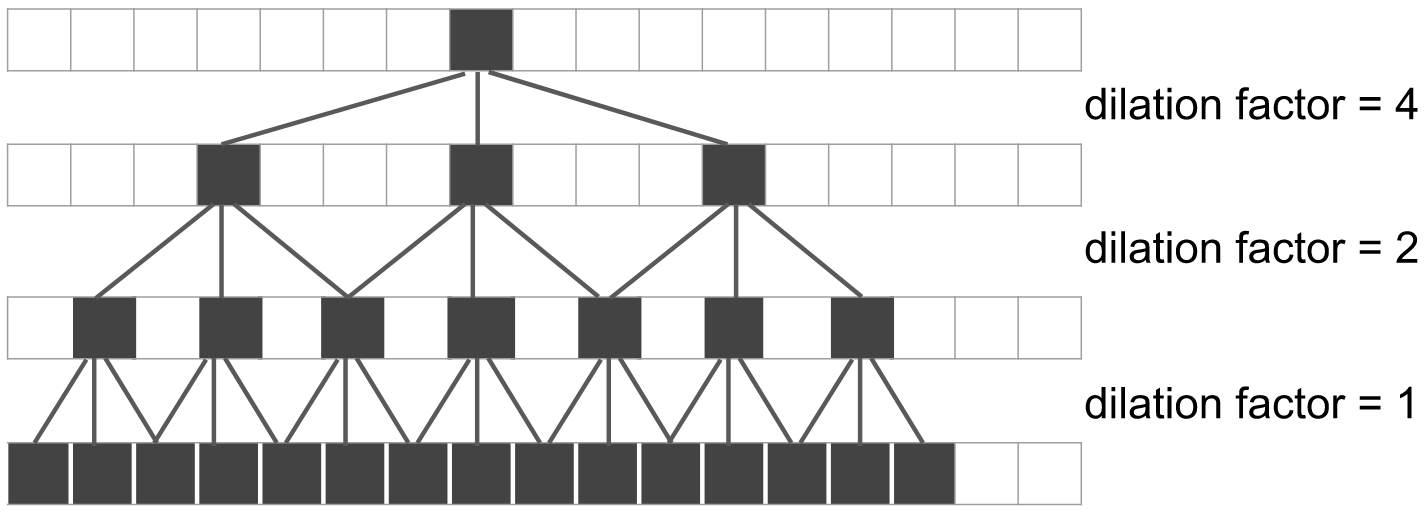}}
\end{minipage}
\caption{Illustration of dilated convolutions, where the dilation
  factors are 1, 2 and 4, respectively. Refer to text for details.
}
\label{fig:dilation}
\end{figure}

\textbf{Dilated Convolution} Dilated convolutions, supporting
exponentially expanding RFs without the loss of resolution or
coverage, can provide a remedy. Also, such expansion can be achieved
without the need to increase the number of parameters.  The basic idea
of dilation is to space out the elements to be summed in convolution
by a dilation factor, as illustrated in Fig.~\ref{fig:dilation}.  The
convolutions in the bottom layer are regular $3 \times 1$
convolutions. The middle layer has a dilation factor of 2, so the
effective RF at each neuron covers 7 audio samples. The top layer
convolutions are dilated by 4, producing a $15 \times 1$
RF/coverage. In general, the receptive filed $R_k$ of a neuron on a
dilated convolution layer $l_k$ is enlarged to:
\begin{equation} \label{eq:dilated}
  R_k=R_{k-1}+((f_k-1)\times d_k\times\prod_{i=1}^{k-1}s_i)
\end{equation}
where $d_i$ is the dilation factor of layer
$l_i$. Comparing with the standard version, dilated convolutions
increase RFs without introducing more parameters.
In addition, dilated convolutions produce exponentially expanding RFs
with depth, which is in contrast to linear expansion produced by
standard convolutions.

\begin{figure}[htb]
\begin{minipage}[b]{1.0\linewidth}
  \centering
  \centerline{\includegraphics[width =  8.5cm]{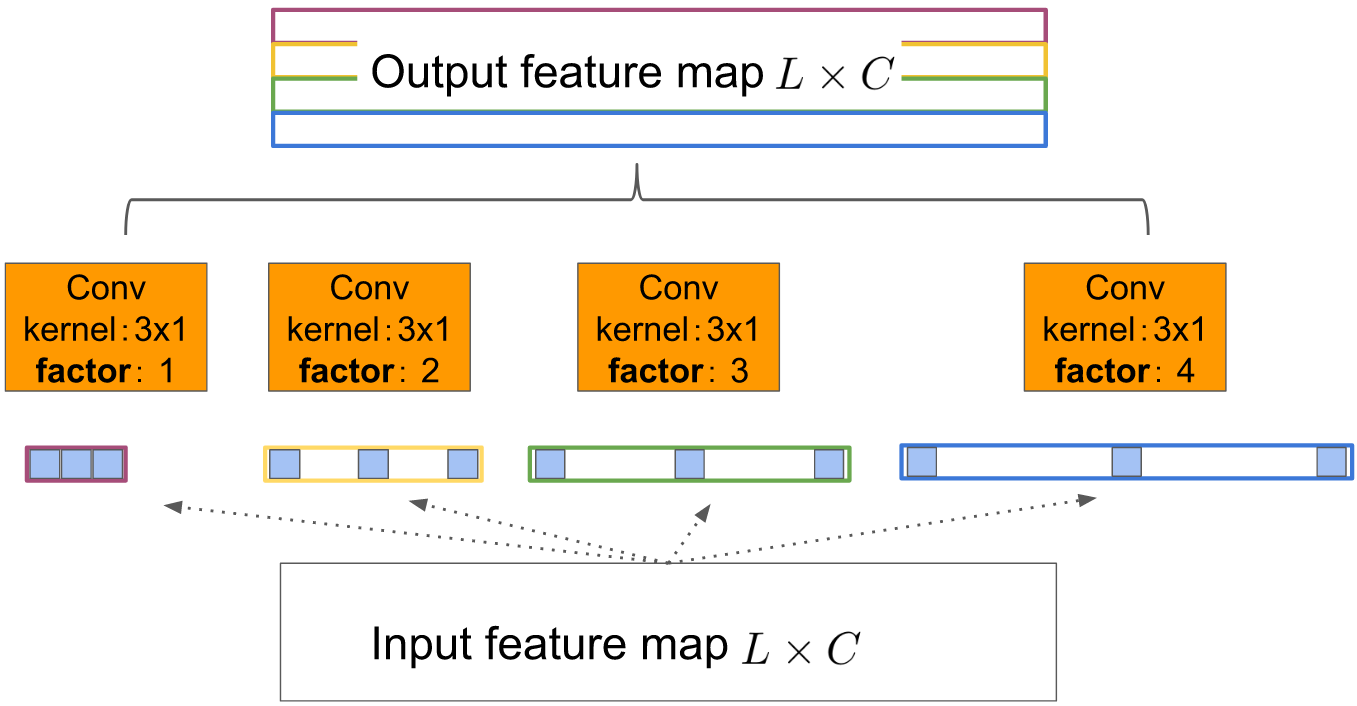}}
\end{minipage}
\caption{Illustration of fusion of multiscale dilation through
  ASPP. Refer to text for more details. }
\label{fig:fuse}
\end{figure}

\textbf{Fusion of multiscale dilations through ASPP}
Dilated convolutions allow us to listen longer.  To extract audio
patterns with great varieties, however, multiple dilation factors
should be involved. Fig.~\ref{fig:fuse} shows an example of applying
dilation convolutions with 4 different factors. How to integrate the
extracted features is a practical issue. In this work, we adopt a
strategy similar to the Atrous Spatial Pyramid Pooling (ASPP) scheme
proposed in DeepLab \cite{chen2018deeplab}. More specifically, we
conduct dilation convolutions of 4 different factors in parallel and
concatenate the resulted feature maps into
outputs. 
These 4 filters are called a {\it dilation group}. Within each group,
the filters have the same number of parameters (3 in the example of
Fig.~\ref{fig:fuse}), but cover different signal ranges because of the
varying filter lengths. For an input feature map of dimension
$L \times C$, where $L$ is the size of the map and $C$ is the number
of channels,
we set the number of the dilation groups to $C/4$, to ensure the
output feature maps to maintain the dimension of $L \times
C$. Comparing with directly utilizing $C$ filters in the standard
fashion, our dilated setup does not increase the number of
parameters, but enables the network to capture longer signals with
varying lengths.

It should be noted that dilation has been utilized in a very recent
work by Tan {\it et al.}  \cite{Tan_2019_T} in a CNN+RNN model. To the
best of our knowledge, our work is the first attempt to explore
dilated convolution to improve FCN models for speech enhancement.

\subsection{Locations to add the dilation module}
To exert the power of dilations + ASPP, the next step is to integrate
the proposed dilation module into our baseline FCN model. To set up a
proper ASPP installation, we run into two practical issues. The first
question is regarding how to set the dilation factors.  Our answer is
based on the fact that the audio patterns to be captured in this work
are mainly human phonetic symbols, whose lengths and frequencies tend
to concentrate around certain ranges. It is necessary to have
dilations of different ratios, but the coverages do not have to be
dramatically diverging in scale. With this observation, we choose a
relatively slow growing sequence, 1, 2, 3 and 4, as the dilation
factors of our ASPP convolutions

\begin{figure}[htb]
\begin{minipage}[h]{1.0\linewidth}
  \centering
  \centerline{\includegraphics[width = 8.5cm]{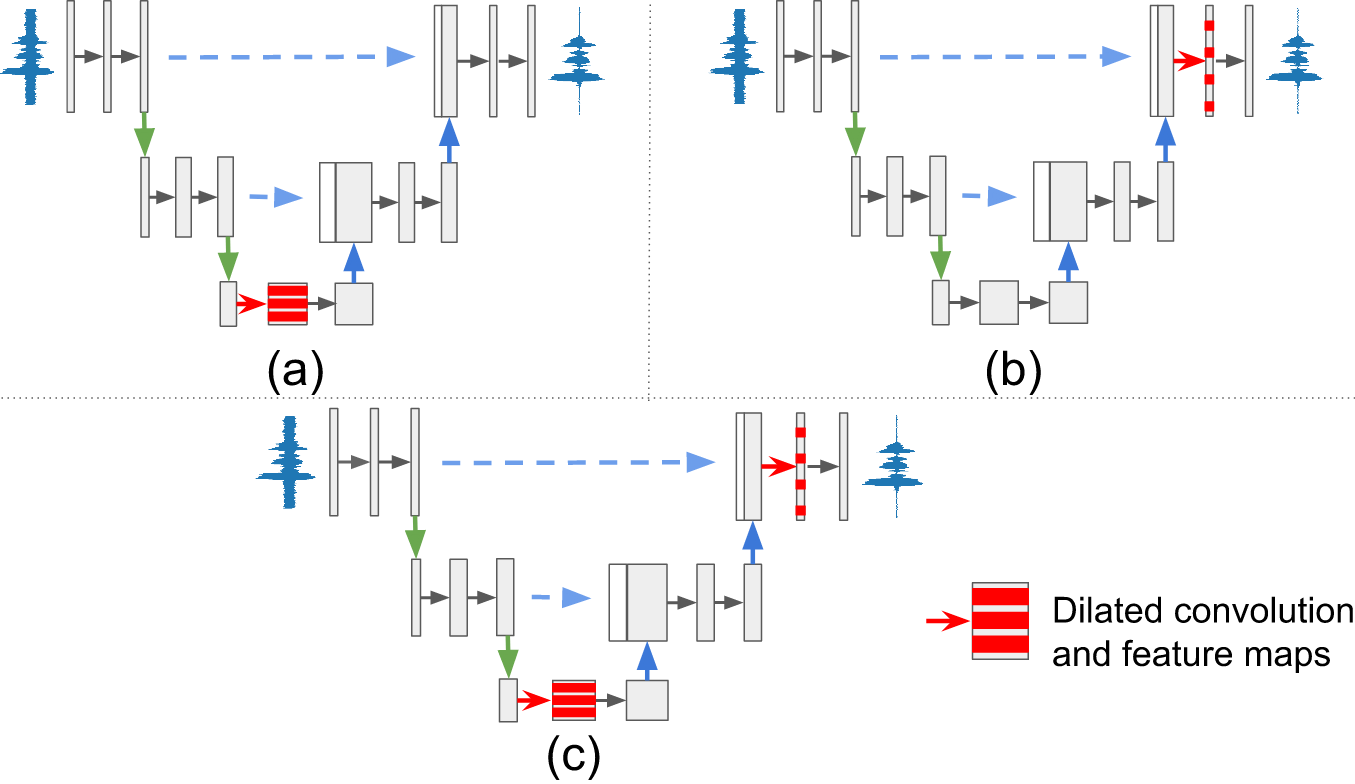}}
\end{minipage}
\caption{Three ASPP replacement schemes: a) adding ASPP in the middle
  layer of the network; b) adding ASPP around the end of the decoding
  path; c) combination of a) and b).}
\label{fig:res-u-net}
\end{figure}

The second question is where to install the ASPP replacement.
While ASPP can be installed anywhere in the network, we hope our
dilated convolutions filters, even with limited number of parameters,
can cover relatively long signal spans. In this regard, the end of the
encoding path would be an ideal 
place to add ASPP, as the neurons on this layer have the largest RFs
in the entire network. The further enlarged RFs by ASPP would provide
a best realization of our goal of ``listening longer''. This setup is
illustrate in Fig.~\ref{fig:res-u-net}.(a).

An alternative place to explore the ASPP replacement could be around
the end of the decoding path, which consists of two convolution
layers, as shown in Fig.~\ref{fig:res-u-net}(b). Replacing the first
of the two layers with an ASPP would potentially allow the network to
decompose the features into different scales, before they will be
finally merged to face the ground-truth.  Adding ASPP here would
provide a (last) chance for the network to {\it correct} the
integration from previous layers.  With the analysis of these two
choices, very naturally, another setup worth exploring would be the
combination of the both.  In our experiments, all three ASPP
replacement schemes, as illustrated in Fig.~\ref{fig:res-u-net}, are
examined.

\vspace{-0.15in}
\section{Experiments and Results} 

{\bf{Data sets}} To evaluate the effectiveness of the proposed
dilation + ASPP for speech enhancement, we conduct experiments on two
groups of datasets.  The first experiment is based on the Noisy VCTK
dataset by Valentini {\it et al.} \cite{valentini2016investigating,
  valentini2017noisy}, which consists of two training sets (28 and 56
speakers respectively) and a test set. We choose the 28-speaker set,
in which 14 males and 14 females were recorded with around 400
sentences for each person.  Noise of 10 different types, either
synthetic or real, have been added to the clean speech with 4
signal-to-noise (SNR) levels (15 dB, 10 dB, 5 dB and 0 dB,
respectively).  Totally there are 11,571 sentences in the training
set. The test dataset contains 824 sentences from two new speakers
(one male and one female). Five types of noise, which are different
from the 10 types in the training set, have been added.  The SNR
values for test set are 17.5 dB, 12.5 dB, 7.5 dB and 2.5 dB,
respectively.  All audio clips are sampled at 48kHz, with each time
point represented as a 24-bit integer.

The second experiment was based on AzBio English sentences that were
developed by Spahr {\it et al.} \cite{spahr2012development}.  The
dataset consisted of 33 lists with 20 sentences in each list. The
sentences ranged from 3 to 12 words (median = 7) in length.  All
speech sentences were spoken by 2 male and 2 female adult speakers,
sampled at 22,050 Hz \cite{spahr2012development}. Two types of masking
noise were added to the sentences to achieve the desired SNRs (3 dB
and 6 dB): speech-spectrum-shaped noise (SSN) and two talker babble
(TTB).

\textbf{Preprocessing} In both experiments, we apply a preprocessing
step to downsample all audio clips to 16kHz, and scale their
amplitudes to [0,1]. The training set are then split randomly into
training, validation and test sets with the ratio of 8:1:1. Each audio
sentence is cut into multiple clips of 1 second long, which are taken
as the inputs to the network. We extract the clips with a half second
overlap as an approach of data augmentation.  End-of-sentence clips,
if short than 0.5 second, are discarded and not included as training
samples.

\textbf {Evaluation metrics} Four evaluation metrics are used in this
work. They are signal-to-noise ratio (SNR), segmental signal-to-noise
ratio (SSNR), perceptual evaluation of speech quality (PESQ) and
short-time objective intelligibility measure (STOI).  SSNR calculates
the average SNRs of short segments (15 to 20 ms long).  PESQ evaluate
the speech quality using the wide-band version recommended in ITU-T
P.862.2 \cite{rec2005p}.  STOI \cite{taal2010short} produces
indicators for the average intelligibility of the degraded speech.

\subsection{Results}

Totally four models are evaluated in our experiments with the Noisy
VCTK dataset. They are the baseline model Speech-U-Net, and three
dilation models with the ASPP replacements at the middle of
Speech-U-Net (bottom layer), end of the network and both locations,
respectively. The evaluations were conducted on two test sets: the
first one is the held-out set, which is the $10\%$ of the training
set; the other is the official test available in the dataset.

\begin{table}[h]
  \caption{Experimental results on Noisy VCTK dataset}
\vspace{0.05in}
\centering
\scalebox{0.8}{
\begin{tabular}{c|c|cccc}
  \hline
  \hline

{\textbf{Dataset}} &
{\textbf{Model}} &
{\textbf{SNR}} &
{\textbf{SSNR}} &
{\textbf{PESQ}} &
{\textbf{STOI}} \\

\hlineB{2.5}
\multirow{4}{*}{Held-out Test}
&{\text{Input}}& 6.040&-0.092&1.467&0.838\\  \cline{2-6}
 &{\text{Speech-U-Net}}& 14.504&7.159&1.849&0.857\\  \cline{2-6}
 &{\text{ASPP-middle}}&\bf{15.042}&\bf{7.718}&\bf{1.882}&0.859\\  \cline{2-6}
 &{\text{ASPP-end}}&14.389&7.181&1.827&0.873\\  \cline{2-6}
 &{\text{ASPP-middle+end}}&14.92&7.665&1.788&\bf{0.877}\\ \hlineB{2.5}

 \multirow{4}{*}{Official Test}
 &{\text{Input}}&8.544&1.878&1.982&0.922\\  \cline{2-6}
 &{\text{Speech-U-Net}}& 17.454&7.955&2.338&0.900\\  \cline{2-6}
 &{\text{ASPP-middle}}&\bf{18.418}&\bf{8.862}&\bf{2.361}&0.902\\  \cline{2-6}
 &{\text{ASPP-end}}& 17.001&8.470&2.262&\bf{0.930}\\  \cline{2-6}
 &{\text{ASPP-middle+end}}&17.529&6.521&2.152&0.928\\ \hline
 \hline
 \hlineB{1.5}
{Official Test}&{\text{SEGAN}}& &7.73&2.16& \\  \cline{1-6}

\end{tabular}
}
\label{T:table1}
\end{table}

\begin{figure}[t]
\begin{minipage}[b]{1.0\linewidth}
  \centering
  \centerline{\includegraphics[width = 8cm]{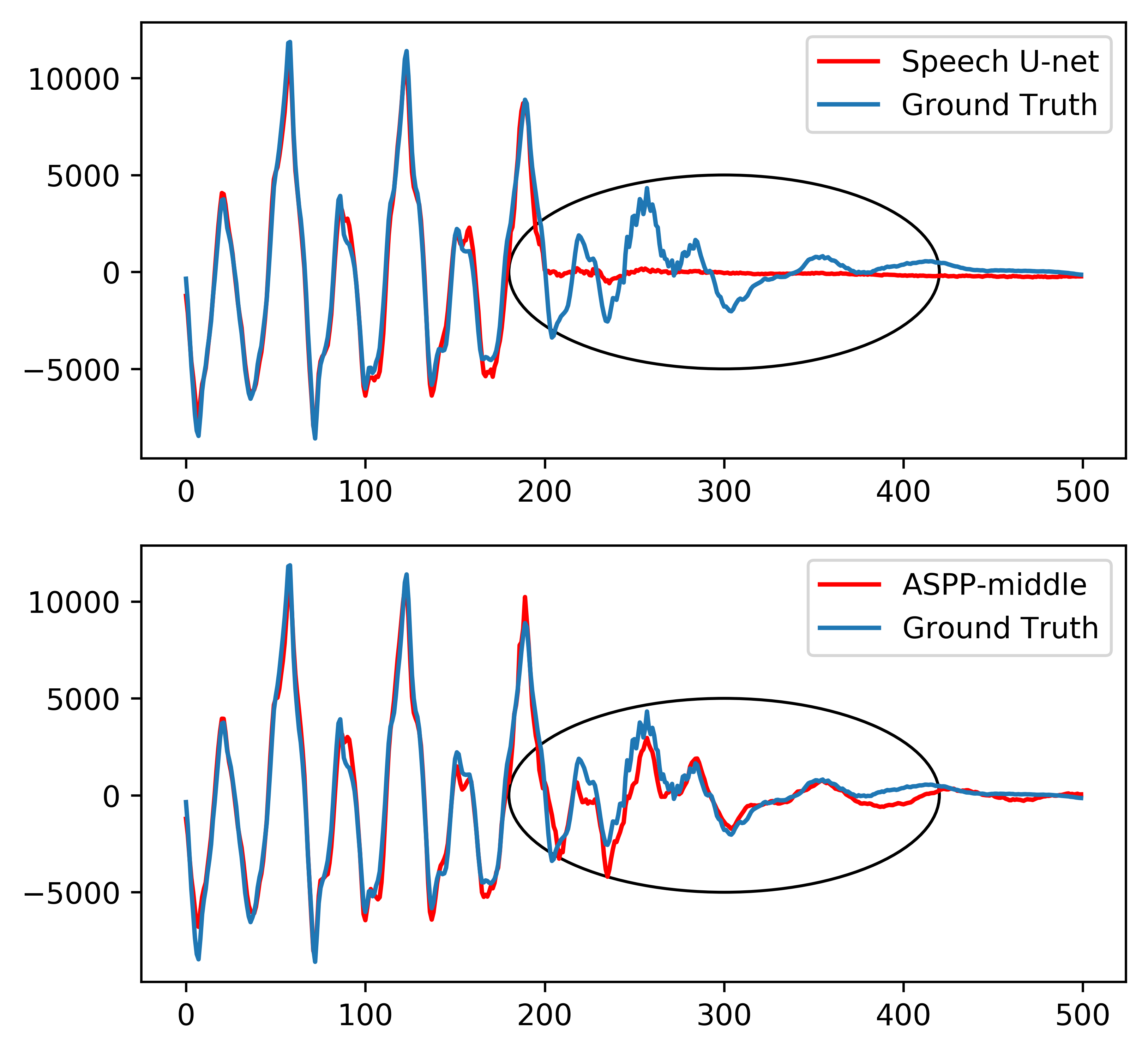}}
\end{minipage}
\caption{Comparison of Speech-U-Net (top row) and ASPP-middle (bottom
  row) on a particular audio clip.}
\label{fig:compare}
\end{figure}

The results are shown in Table~\ref{T:table1}. It is evident that the
baseline Speech-U-Net already performs rather impressively, achieving
an average enhancing performance of 9 dB. Among the three dilation
models, ASPP replacement at the middle (ASPP-middle) produces the best
results, significantly outperforming all other competing solution in
SNR and SSNR.  ASPP-end model does not help, performing even worse
than the baseline model in SNR and SSNR. The combination of {\it
  at-middle} and {\it at-end}, unsurprisingly, has performance
in-between of the two installations. Comparing the results on the
held-out and official test sets, the improvements made by ASPP-middle
are more significant for the latter (official test set). Considering
that the official test data are acquired from new speakers with
different SNRs, therefore have different data distributions than the
training set, the comparison indicates that the ASPP-middle is more
robust and has a better generalization capability than the baseline
model. Such desired properties should be attributed to the enhanced
multi-scale processing brought by the dilation operations. In other
words, ``listening longer'' does make the network hear
better. Fig.~\ref{fig:compare} shows the comparison of ASPP-middle
with Speech-U-Net on a particular audio segment. Ground-truth
waveforms are shown in blue, and red lines are the predictions. For
the highlighted audio segment in the waveform
pictures, 
Speech-U-Net makes rather flat predictions, failing to capture the
fluctuations. Our ASPP-middle, on the other handle, makes accurate
predictions for the entire segment.

For PESQ and STOI, none of the three ASPP models produces improvements
over the baseline.  This can be in part explained by the choice of the
objective function in our models. The network updates in our models
are driven to minimize the $L_1$ difference between predictions and
ground-truth, which is highly related to SNR/SSNR, but does not
directly involve perception and intelligibility components.  To
replace the $L_1$ distance with a STOI-based objective
\cite{Tsao_2018_T}, our models are expected to produce improved
performance measured by PESQ and/or STOI.

It should be noted that our ASPP-middle model also has a higher
average SSNR value than the reported number from SEGAN
\cite{pascual2017segan}, a state-of-the-art speech enhancement
solution. While we do not intend to make a head-to-head quantitative
comparison, as different experiment setups are used in the studies,
the significant improvements in SSNR can nevertheless be regarded as a
side evidence of the effectiveness of our ASPP-middle model. In
addition, our ASPP-middle network, working as a generator, can be
connected with a discriminator to make a full GAN model for further
improvements.


Based on the results and observation from the VCTK experiment, we
chose ASPP-middle as our proposed model. We further evaluate its
capability using the AzBio sentence dataset. Speech-U-Net is still
taken as the baseline model. The comparisons are shown in
Table~\ref{T:DrXu}. Similar to the first experiment, ASPP-middle
produces larger speech enhancements than the baseline, measured in SNR
and SSNR. In summary, our proposed ASPP-middle approach consistently
improves over the baseline, demonstrating the benefits of dilation
convolutions, as well as the fusion and installation setups we
designed.


\begin{table}
  \caption{Experimental results of Speech-U-Net and ASPP-middle on AzBio dataset}
  \vspace{0.05in} \centering \scalebox{0.9}{
\begin{tabular}{c|cccc}
  \hline
  \hline

  {\textbf{Model}} &
                     {\textbf{SNR}} &
                                      {\textbf{SSNR}} &
                                                        {\textbf{PESQ}} &
                                                                          {\textbf{STOI}} \\

\hlineB{2.5}
  {\text{Input}}&2.697&-4.204&1.062&0.816\\  \cline{1-5}
  {\text{Speech-U-Net}}&9.091&-1.534&\bf{1.334}&\bf{0.834}\\  \cline{1-5}
  {\text{ASPP-middle}}&\bf{9.348}&\bf{-1.369}&1.315&0.797\\  \cline{1-5}

\end{tabular}
}
\label{T:DrXu}
\end{table}

\section{Conclusions}

The performance of Speech-U-Net indicates that FCN is a successful
architecture for waveform-based SE. The ASPP module with dilated
convolution expands the RFs of the network when mounted onto a proper
place of FCN. Meanwhile, it does not introduce extra parameters to the
network. These advantages lead to improved achievements in both
datasets we examined. We are currently working to link ASPP
modules 
to develop FCN + RNN-based \cite{chen_mlmi_2017, wang_isbi_2019}
solutions for short delay and quick response, as well as to explore
the integration with other deep neural networks, e.g., GAN models.

\newpage 
\small
\bibliographystyle{IEEEtran}
\bibliography{waspaa_clean}

\end{document}